# Rydberg exciton states and near-infrared light-emitting diode in monolayer MoTe$_2$ devices

*Sebastian Yepez Rodriguez[1], Marshall A. Campbell[1], Jinyu Liu[1], Luis A. Jauregui[1,*]*

[1]Department of Physics and Astronomy, University of California, Irvine, CA 92697 USA

ABSTRACT: Excitons, or bound electron-hole pairs, play a crucial role in the optical response of monolayer, 2H-phase transition-metal dichalcogenides (TMDs). They hold significant promise for the development of novel quantum opto-electronic devices due to their large binding energies and strong spin-orbit coupling. Among the monolayer TMDs, MoTe$_2$ stands out because of its bandgap in the near-infrared (NIR) regime. Here, we report the experimental observation of NIR Rydberg excitons and conduction band-split charged excitons, in high-quality, boron nitride (BN)-encapsulated monolayer MoTe$_2$ devices, probed by photoluminescence and electroluminescence spectroscopy. By employing a graphite bottom gate, we successfully modulate the emission intensity of various excitonic species. Additionally, our device fabrication process within an argon-filled glove box ensures clean TMD/metal electrode interfaces, enabling the construction of p-n junctions near the electrodes. Our work significantly advances our understanding of excitons in monolayer TMDs and contributes to the application of $MoTe_2$ in NIR quantum opto-electronic devices.

Monolayers of transition metal dichalcogenides (TMDs) in the 2H-phase are direct bandgap semiconductors with a two-dimensional (2D) nature. In these systems, the dielectric screening of carriers is significantly reduced, leading to enhanced Coulomb interactions.[1,2] The optical properties of monolayer TMDs are governed by excitons, which are bound electron-hole pairs. Thanks to the enhanced Coulomb interactions, these excitons have large binding energies that result in strong optical resonances.[3–5] To date, much interest has been focused on monolayer TMDs whose electronic bandgap lie in the visible regime of the electromagnetic spectrum such as $WS_2$, $WSe_2$, $MoS_2$ and $MoSe_2$.[6–11] On the other hand, $MoTe_2$ has a bandgap in the near-infrared (NIR) regime[12,13] very similar to that of silicon, which makes it a promising candidate to be used in NIR opto-electronic devices.[14–16] The bandgap of $MoTe_2$ can be tuned through electrostatic gating[17–19] and tensile strain,[20–24] and its crystal structure may undergo a phase transition from 2H phase to 1T' phase when enough gating or strain is applied.[17–24] As the quality of monolayer samples rapidly increases, excited states of excitons that resemble the hydrogenic Rydberg series have recently been observed in photoluminescence (PL) measurements on monolayer $MoTe_2$, using a Salisbury-screen geometry.[25] This finding makes $MoTe_2$ a good candidate to be used in NIR quantum sensing, given the larger Bohr radii of the excited excitons in the Rydberg series.[26–28]

Monolayer TMDs exhibit strong spin-orbit coupling (SOC), and it is particularly high in $MoTe_2$ due to the presence of heavy tellurium atoms.[29] The strong SOC allows for the emergence of chiral topological excitons and polaritons through the application of position-dependent magnetic fields,[30] or through microcavities or surface acoustic waves in combination with a constant magnetic field.[31] Due to broken inversion symmetry in 2D TMDs, the chiral nature of excitons in monolayer TMDs can be probed using circularly polarized light of opposite chirality, which couple to the different energy-degenerated valleys, K and K', in momentum space. It has also been shown that the polarization of chiral valley excitons can be generated and controlled by in-plane electric fields.[32] The electrical generation of excitons[33] or electroluminescence (EL) is an important capability of monolayer TMDs, useful for studying fundamental physics phenomena and opto-electronic device applications. Particularly at cryogenic temperatures, it has been shown that electrically generated excitonic emission comes from localized or charged excitons, in addition to neutral excitons.[32–37]

In the present work, we report the experimental observation of neutral, charged, and excited exciton species in monolayer 2H-MoTe$_2$, probed by both photoluminescence and electroluminescence measurements at low temperature ($T$ = 4 K). The extremely high quality of our MoTe$_2$ samples allows us to identify the observed optical resonances as the Rydberg series of excitons, as well as to resolve the fine structure of the charged ground exciton state, using a simple device geometry. Our observation and analysis of the various excitonic species revealed by both PL and EL in monolayer MoTe$_2$ can lay the foundations to create novel quantum devices based on 2D semiconductors with a bandgap similar to that of silicon, where the role of strong SOC can be enhanced and facilitate the creation of topological opto-electronic devices.

We fabricated a BN-encapsulated monolayer MoTe$_2$ heterostructure (Fig. 1a, b) using the dry-transfer technique. The high-quality bulk single crystals of 2H-MoTe$_2$ were grown using the flux method, with excess tellurium serving as flux. Details about crystal growth are in the Supporting Information (SI). The mechanical exfoliation and flake transferring processes were performed inside an argon-filled glovebox to avoid any exposure of the MoTe$_2$ monolayer samples to air (SI

Sec. 2). The monolayer nature of our MoTe$_2$ sample was initially determined by the contrast under an optical microscope and then confirmed by Raman spectroscopy at low temperature ($T$ = 4 K), as shown in Fig. 1c. The Raman spectra, under illumination with a 532 nm continuous wave laser, reveals distinct features between the monolayer and few-layer regions. The absence of the B$_{2g}$ mode and the weaker Raman response of the E$^1_{2g}$ mode in the monolayer region,[12,38] compared to the few-layer region, indicate that our sample is indeed a single layer of MoTe$_2$. Moreover, the strong Raman response of the A$_{1g}$ mode in the monolayer region further confirms this identification.[38]

Next, we measured the PL spectrum of our monolayer MoTe$_2$ heterostructure at $T$ = 4 K (Fig. 1d) and observed three different peaks with energies in the NIR regime. We attribute these peaks to the ground state neutral exciton, $X^0_{1s}$, the ground state charged exciton, $X^T_{1s}$, and first excited state exciton, $X^0_{2s}$, of the Rydberg series, in accordance with previous reports on MoTe$_2$ monolayers.[25,39] Fig. 1d shows two important features of the 1s state probed by PL: the exciton peak has a narrow linewidth (~4 meV) and the $X^T_{1s}/X^0_{1s}$ intensity ratio is fairly small (~13 %), both of which are indicative of high sample quality.[25] Moreover, the observation of $X^0_{2s}$ without the use of PL enhancement[25] further confirms the high quality of our MoTe$_2$ sample.[39] In the 2D Rydberg series model for the excitonic quantum states in monolayer TMDs, the exciton emission energies follow the relation[40] $E_n = E_g - E_{b(1)}/(2n-1)^2$, where $E_g$ is the free-particle bandgap, $E_{b(1)}$ is the binding energy of the ground state, and $n = 1,2,3,...$ is the quantum number of the Rydberg energy states (1s, 2s,...). The binding energy of the Rydberg states reads $E_{b(n)} = 2\hbar^2/(\mu r_n^2)$, where $\hbar$ is the Planck's constant, $\mu$ is the exciton reduced mass, and $r_n$ is the exciton effective Bohr radius. Thus, the binding energy of $X^0_{2s}$ is expected to be nine times smaller than that of $X^0_{1s}$, making $r_2$ three times larger than $r_1$. Based on the emission energies of $X^0_{1s}$ and $X^0_{2s}$ (Fig. 1d), we estimated the binding energies of the 1s and 2s excitons to be around 151 meV and 17 meV, respectively, which agrees well with the expected binding energy ratio of $X_{1s}^0/X_{2s}^0 \sim 9$. These results indicate that the Bohr radius of $X_{2s}^0$ is about three times larger than that of $X_{1s}^0$ in our sample. The approximate size of the Rydberg excitons in MoTe$_2$ is illustrated in the inset of Fig. 1d, where the 2s exciton corresponds to the large sphere and the 1s exciton to the small sphere. Both 1s and 2s excitons emit NIR light when electrons and holes recombine radiatively. We also estimated the free-particle energy bandgap to be 1.343 eV, which agrees well (7% difference) with the bandgap obtained elsewhere for BN-encapsulated monolayer MoTe$_2$ through magneto-optical spectroscopy.[41]

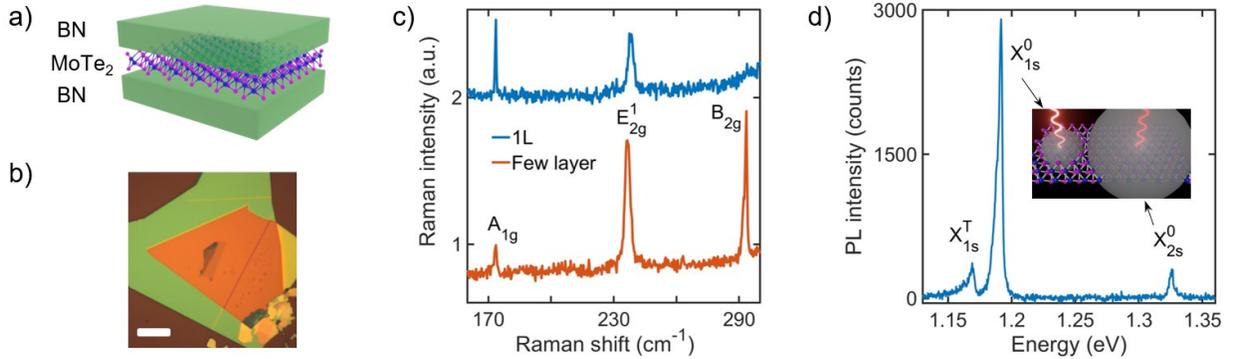

**Figure 1.** Rydberg exciton states in BN-encapsulated monolayer MoTe$_2$. a) Schematic of the BN-encapsulated monolayer MoTe$_2$ heterostructure. b) Optical image of the heterostructure. Scale bar: 20 μm. c) Raman spectrum of monolayer and few-layer regions of the MoTe$_2$ sample. $A_{1g}$, $E^1_{2g}$ and $B_{2g}$ correspond to different vibrational modes of 2H-phase MoTe$_2$. d) Photoluminescence spectrum of monolayer MoTe$_2$ showing the 1s neutral exciton, $X_{1s}^0$, 1s charged exciton, $X_{1s}^T$, and 2s neutral exciton, $X_{2s}^0$, of the Rydberg series. Inset: Illustration of 1s and 2s excitons, corresponding to smaller and larger Bohr radii, respectively.

Rydberg excitons in monolayer TMDs are very sensitive to changes in their electronic landscape[11,25,42,43], so we fabricated another BN-encapsulated monolayer MoTe$_2$ heterostructure with a 7 nm-thick graphite at the bottom (Fig. 2a, b) to apply a gate voltage, $V_G$. In this manner, we were able to tune the carrier density of monolayer MoTe$_2$. Fig. 2c, e show colormaps representing the gate voltage dependence of the PL intensity for the 1s and 2s states, respectively. The peak intensity of the 1s neutral exciton indicating the charge neutrality point (CNP), happens around $V_G$ = -0.1 V. Fig. 2d, f are line cuts of Fig. 2c, e showing that, for both the 1s and 2s states, the PL emission is dominated by neutral excitons when the chemical potential is close to the CNP at $V_G$ = 0 V, and then it becomes dominated by charged excitons when $V_G$ = 1 V and $V_G$ = -1 V, away from CNP. This is the expected behavior of Rydberg excitons in monolayer TMDs upon an increase of the carrier density,[11,25,42,43] since there are more free carriers that prefer to bind to excitons to lower the total energy of the 2D system.

We used a Fast-Fourier transform (FFT) filter for the 2s state PL data in Fig. 2e, f, in order to minimize the spectral etaloning interference effect, which is intrinsic to the optical sensor in our PL setup (SI Sec. 4). Taking a closer look at the energies of neutral and charged excitons of the 1s state (Fig. 2c, d), we note there are in total three main peaks: two bright charged exciton peaks, $X_{1s}^{T,1}$ at 1.141 eV and $X_{1s}^{T,2}$ at 1.149 eV, arising mainly from SOC splitting of the conduction band,[44] and the neutral exciton peak, $X_{1s}^{0}$, at 1.172 eV.

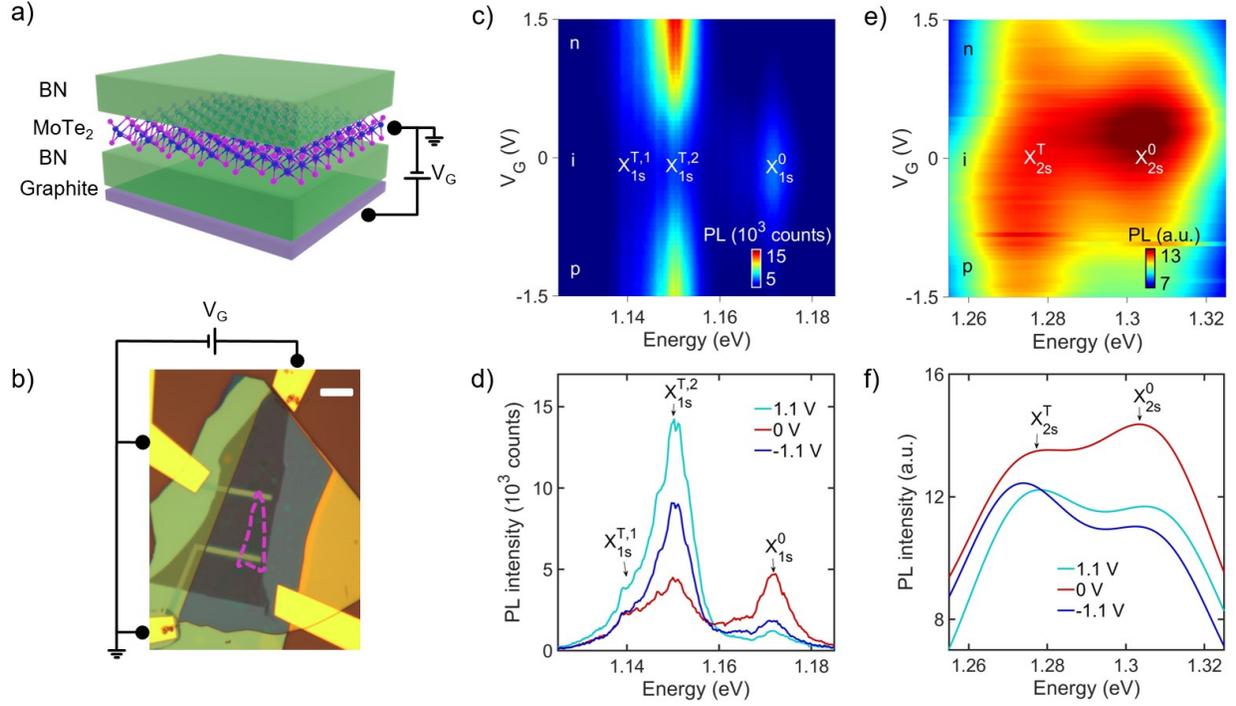

**Figure 2.** Gate voltage dependence of Rydberg states in monolayer MoTe$_2$. a) Schematic of the BN-encapsulated monolayer MoTe$_2$ heterostructure with bottom graphite gate. A gate voltage $V_G$ can be applied between the monolayer MoTe$_2$ sample and the graphite gate. b) Optical image of the heterostructure. Sample is outlined in pink. $V_G$ is the same as in (a). Scale bar: 10 µm. c) PL intensity colormap as a function of energy and $V_G$ for the 1s Rydberg state, showing two charged 1s excitons $X_{1s}^{T,1}$ and $X_{1s}^{T,2}$, and the neutral 1s exciton $X_{1s}^0$. d) Line cuts of (c) at $V_G$ = 1.1 V, 0 V and -1.1 V. e) PL intensity colormap as a function of energy and $V_G$ for the 2s Rydberg state, showing the charged 2s exciton $X_{2s}^T$ and the neutral 2s exciton $X_{2s}^0$. f) Line cuts of (e) at $V_G$ = 1.1 V, 0 V and -1.1 V. In (c) and (e) the letters n, i and p denote electron-doped, intrinsic, and hole-doped regimes, respectively.

To confirm our identification of the 1s and 2s neutral and charged exciton states in the PL spectra of monolayer MoTe$_2$, we performed power-dependent PL measurements, as shown in Fig. 3a, c. Focusing first on the 1s state, we fitted a sum of four Lorentzian functions to the PL spectra and calculated the integrated PL intensity of the four excitonic species as power increases (Fig. 3b). Included in Fig. 3b are power-law fittings to the integrated PL intensity of the form
7

$I_{PL} = CP^{\kappa\alpha}$, where $C$ and $\kappa$ are dimensionless constants, $P$ is the laser power, and $\alpha$ is a coefficient indicating the nature of the excitonic species. We calculated $\kappa = 0.67$ by setting $\alpha \equiv 1$ for the neutral exciton $X_{1s}^0$ and used this as a reference to extract the $\alpha$ coefficients for the other excitonic species.[37] We used the constant $\kappa$ in the power-law fittings to account for the nonlinear increase of $X_{1s}^0$ intensity with power, which is likely a result of loss of exciton population to a defect-related localized state[45] near $X_{1s}^0$ (details about this additional peak in SI Sec. 5). In this scenario, both charged excitons $X_{1s}^{T,1}$ and $X_{1s}^{T,2}$ increase superlinearly with power ($\alpha = 1.35$ for $X_{1s}^{T,1}$ and $\alpha = 1.28$ for $X_{1s}^{T,2}$), in accordance with previous studies.[37,46,47]

In the case of 2s state excitations, we fitted a sum of two Lorentzian functions to the PL spectra accounting for the 2s state neutral exciton $X_{2s}^0$ and charged exciton $X_{2s}^T$. Fig. 3d shows the increase rate of the integrated PL intensity for $X_{2s}^0$ and $X_{2s}^T$ with power. From power-law fittings, again setting $\alpha \equiv 1$ for $X_{2s}^0$, we note that $X_{2s}^T$ increases sublinearly ($\alpha = 0.88$) with power rather than superlinearly like $X_{1s}^{T,1}$ and $X_{1s}^{T,2}$ do. This trend may be related to the larger linewidths of 2s exciton species (~40 meV) compared to those of 1s exciton species (~12 meV). The large linewidths of the 2s excitations could have prevented us from resolving peaks associated with localized states energetically near $X_{2s}^0$ and $X_{2s}^T$. Loss of exciton population to localized states[45] would then affect the increase of $X_{2s}^T$ intensity with power.

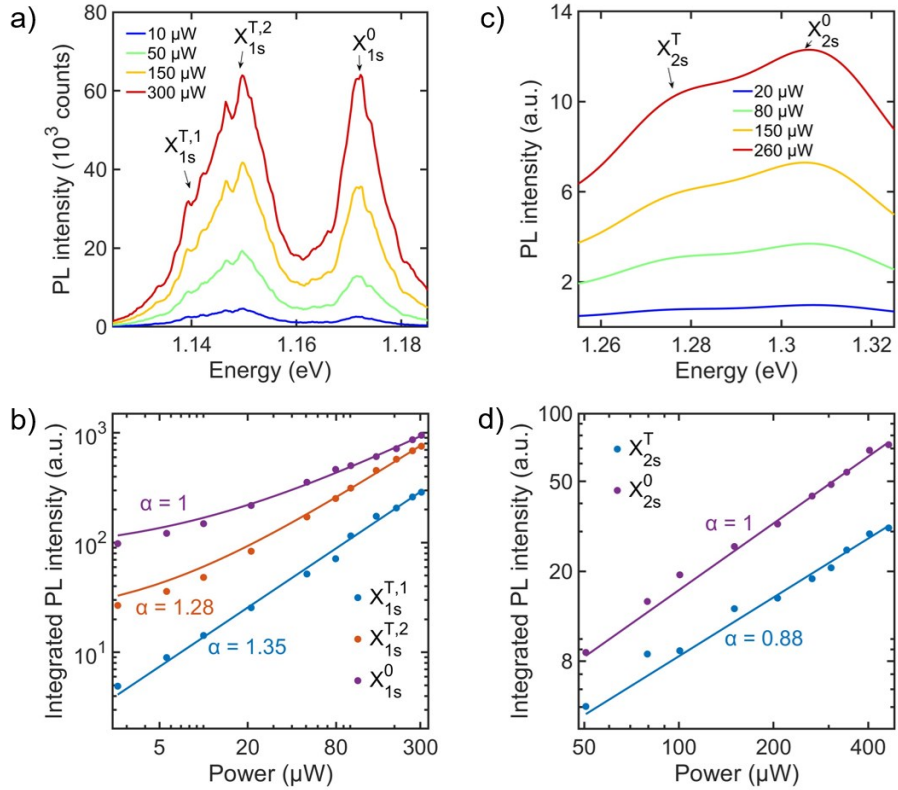

**Figure 3.** Laser power dependence of Rydberg states. a) and c) PL spectra at different laser power for the 1s and 2s states, respectively. b) and d) Integrated PL intensity as a function of power extracted from Lorentzian fittings to the power-dependent PL spectra for the 1s and 2s states, respectively. The solid lines are power law fittings of the integrated PL intensity dependence on power. The coefficients $\alpha$ are extracted from these fittings and indicate the nature of the excitonic species.

Neutral and charged excitons can be created in 2D semiconducting TMDs not only by optical excitation but also by electrical injection of carriers. Several methods to generate excitons from electrical injection include p-n junctions from split-gate heterostructures,[34,48] tunneling diodes,[35,49] devices using AC voltage between semiconductor and gate,[36,37] and the use of high or low work function materials to locally form p-n junctions.[50–52] By placing monolayer MoTe$_2$ on top of two gold electrodes (Fig. 2b and Fig. 4a) in an inert environment, therefore keeping a clean TMD/metal interface, we realized p-n junctions near the electrodes by (1) locally hole-doping the sample on top of the electrodes[52] and (2) electron-doping the rest of the sample using the graphite bottom gate. In order to perform EL measurements on our monolayer MoTe$_2$, we applied a bias (source-drain) voltage, $V_{DS}$, between the two gold electrodes underneath each end of the sample to create an in-plane electric field so electrical current can flow through. Fig. 4a shows a spatial map of the integrated EL intensity after applying a gate voltage $V_G$ = 12 V and bias voltage $V_{DS}$ = 30 V by grounding electrode 1 and applying -30 V to electrode 2. Notably, the EL signal comes mainly from the regions of the sample on top of and surrounding both electrodes. Fig. 4b presents an EL spectrum (red curve) collected at a spot of the sample near electrode 2 applying $V_G$ = 10 V and $V_{DS}$ = 30 V, and compares it to the PL spectrum (blue curve) collected at a spot near the middle of the sample using $V_G = V_{DS} = 0$ V. We identified in this EL spectrum the same three peaks that we observed in the PL spectra, namely, the 1s low energy charged exciton $X_{1s}^{T,1}$ at 1.140 eV, the 1s high energy charged exciton $X_{1s}^{T,2}$ at 1.153 eV, and the 1s neutral exciton $X_{1s}^{0}$ at 1.172 eV.

In order to understand how the excitonic EL signal is generated, we recorded $I_{DS}$-$V_{DS}$ curves at various gate voltages and observed the characteristic behavior of a diode. As shown in Fig. 4c, after $V_{DS}$ has reached a certain threshold voltage, the monolayer MoTe$_2$ sample allows a current of electrons to flow from electrode 2 to electrode 1. As $V_{DS}$ is increased further, the sample enters the "linear regime" of the diode where $I_{DS}$ is approximately linear with $V_{DS}$. In this regime, we estimate the resistance of the monolayer MoTe$_2$ sample to be around ~170 k$\Omega$, including the contact resistance at the two sample/electrode interfaces with $V_G$ = 12 V and $V_{DS}$ = 30 V. Fig. 4c also illustrates that the threshold voltage required to initiate the diode decreases as $V_G$ increases. This is because the device needs a smaller electric field to produce a current when more carriers are present in the sample, thus confirming the diode behavior of our monolayer MoTe$_2$ device.

Fig. 4d schematically shows how the excitonic emission is generated through the p-n junctions that are formed near the electrodes. First, the clean sample/electrode interface allows the gold electrode to locally hole-dope (p) the sample directly on top of it,[52] since the work function of gold (5.31 eV)[53] lies below the valence band maximum of monolayer MoTe$_2$ (5.04 eV).[54] Then, by applying the gate voltage V$_G$, the rest of the sample is electron-doped (n) and p-n junctions in monolayer MoTe$_2$ are created around each electrode. When the bias voltage V$_{DS}$ is applied in a forward configuration and reaches a certain threshold, the energy barrier between n and p regions of the sample starts to get reduced. In other words, the resistance to flow between these regions becomes finite. Hence, the diode is in the "on" state, as opposed to the "off" state when $I_{DS}/V_{DS} = 1/R = 0$, and the resistance $R$ is infinite. Electrons then start to flow towards the p region and holes flow towards the n region, making it possible for pairs of electrons and holes to bind in the depletion region (on top of and surrounding the electrodes) to form excitons. These excitons can then recombine radiatively and produce the EL spectrum observed (Fig 4b). In this way, we realized a monolayer MoTe$_2$ NIR light-emitting diode (LED).

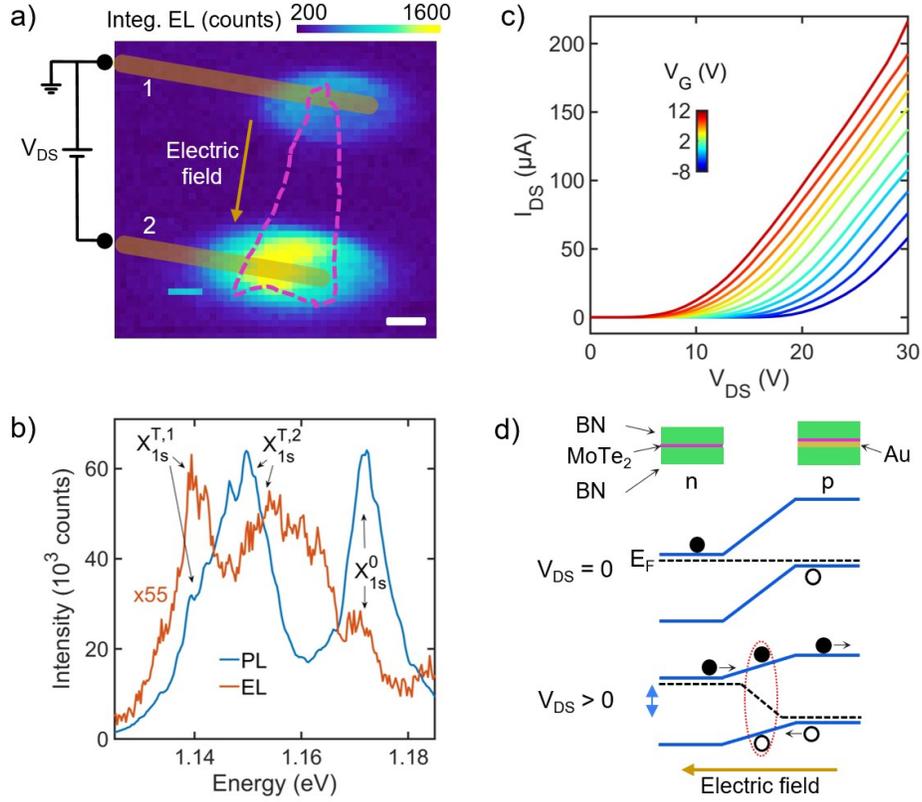

**Figure 4.** Electroluminescence in monolayer MoTe$_2$. a) Spatial map of the integrated EL intensity of monolayer MoTe$_2$ using $V_G$ = 12 V and $V_{DS}$ = 30 V. The sample is outlined in pink. Electrode 1 is grounded and electrode 2 is at -30 V. Scale bar: 4 μm. b) PL spectrum (blue) collected at the center of the sample using an incident power of 300 μW, and EL spectrum (red) collected near electrode 2 in the sample using $V_G$ = 10 V and $V_{DS}$ = 30 V (counts are multiplied by a factor of 55 for the EL spectrum). The neutral exciton $X_{1s}^0$ and charged excitons $X_{1s}^{T,1}$ and $X_{1s}^{T,2}$ are identified in both spectra. c) $I_{DS}$-$V_{DS}$ curves of monolayer MoTe$_2$ device. $V_G$ varies from -8 V to 12 V in steps of 2 V. d) Schematic representation of excitonic emission from p-n junctions around electrodes. Top: the n region is formed in the sample far from the electrode, and the p region is formed in the sample on top of the electrode. Middle and bottom: electronic band schematics in the interface between n and p regions (depletion region) leading to excitonic emission when $V_{DS} > 0$.

In conclusion, we have experimentally observed Rydberg exciton states and successfully realized an NIR light-emitting diode in monolayers of high-quality MoTe$_2$ through PL and EL spectroscopy. Our work adds valuable insights to the exploration of monolayer $MoTe_2$ for its potential application as active material in opto-electronic devices. Furthermore, it contributes to the broader understanding of $MoTe_2$ as a promising candidate for quantum sensing applications.

ASSOCIATED CONTENT

**Supporting Information**. The following files are available free of charge.

Further details of the $MoTe_2$ single crystal growth, materials characterization, device fabrication, Raman, PL, and EL measurements, etaloning effect and FFT filtering, Lorentzian fittings of the luminescence spectra, and power dependence of the EL spectra. (PDF)

AUTHOR INFORMATION

**Corresponding Author**

Luis A. Jauregui – Department of Physics and Astronomy, University of California at Irvine, California 92697, USA.

Email: lajaure1@uci.edu

**Author Contributions**

S.Y.R and L.A.J. designed the experiment. J.L. grew the crystals. S.Y.R. performed the fabrication of the samples and devices and the electrical measurements. S.Y.R. and M.A.C. performed the optical measurements. The manuscript was written through contributions of all authors. All authors have given approval to the final version of the manuscript.

**Acknowledgements**

This research was primarily supported by the National Science Foundation NSF-CAREER (DMR 2146567). This research was partially supported by the National Science Foundation Materials Research Science and Engineering Center program through the UC Irvine Center for Complex and Active Materials (DMR-2011967). M.A.C acknowledges the support from the GEM foundation through a GEM foundation fellowship. The authors acknowledge Michael T. Pettes from Los Alamos National Laboratory for his help with Raman characterization of the single layer sample. The authors also acknowledge Triet Ho for helping with rendering 3D images in Blender.

**Notes**

The authors declare no competing financial interest.